\documentclass[11pt,a4paper]{article}

\usepackage{amsmath,amssymb}
\usepackage{epsfig,graphicx}
\usepackage{subfigure}
\usepackage{graphicx}
\usepackage{rotating}
\usepackage{cancel}
\usepackage{bm}
\usepackage{color}
\usepackage{comment}
\usepackage{psfrag}
\usepackage[backend=biber,citestyle=numeric-comp,natbib=true,sorting=none]{biblatex}

\renewcommand\[{\left[}

\newcommand{\J}[1]{\color{magenta} #1 \color{black}}

\newcommand{\exclude}[1]{}

\def\beq{\begin{equation}}
\def\eeq{\end{equation}}
\newcommand{\C}[1]{\mathcal{#1}}

\usepackage{color}

\usepackage{footmisc,multirow}
\usepackage{subfigure}
\usepackage{amsmath,slashed}

\usepackage{graphicx}
\renewcommand{\d}{{\text{d}}}
\newcommand{\be}{\begin{equation}}
\newcommand{\ee}{\end{equation}}

\newcommand{\bee}{\begin{eqnarray}}
\newcommand{\eee}{\end{eqnarray}}
\newcommand{\beeq}{\begin{equation}}
\newcommand{\eeeq}{\end{equation}}

\renewcommand{\vec}{\bf}

\begin{document}

\title{\vspace{-2cm}{\normalsize  \mbox{}\hfill IPPP/19/10}\\
\vspace{2.5cm} 
\LARGE{A fresh look at ALP searches in fixed target experiments}\vspace*{0.5cm}}

\author{Lucian Harland--Lang$^{1}$, Joerg Jaeckel$^{2}$ and Michael Spannowsky$^{3}$\\[2ex]
\small{\em $^1$Rudolf Peierls Centre, Beecroft Building,} \\
  \small{\em Parks Road, Oxford, OX1 3PU, United Kingdom}\\[0.5ex]  
\small{\em $^2$Institut f\"ur theoretische Physik, Universit\"at Heidelberg,} \\
\small{\em Philosophenweg 16, 69120 Heidelberg, Germany}\\[0.5ex]  
\small{\em $^3$Institute for Particle Physics Phenomenology, Department of Physics,} \\
  \small{\em South Road, Durham DH1 3LE, United Kingdom}
  \\[0.8ex]}

\date{}
\maketitle

\begin{abstract}
\noindent A significant number of high power proton beams are available or will go online in the near future. This provides exciting opportunities for new fixed target experiments and the search for new physics in particular. In this note we will survey these beams and consider their potential to discover new physics in the form of axion-like particles, identifying promising locations and set ups. To achieve this, we present a significantly improved calculation of the production of axion-like particles in the coherent scattering of protons on nuclei, valid for lower ALP masses and/or beam energies. We also provide a new publicly available tool for this process: the \texttt{Alpaca} Monte Carlo generator. This will impact ongoing and planned searches based on this process.  
\end{abstract}

\section{Introduction}
\label{sec:intro}
Very weakly coupled new MeV-GeV scale particles provide an interesting avenue to exploring dark sectors which, perhaps most importantly, may contain dark matter itself.
Fixed target experiments offer exciting possibilities to search for these increasingly popular particles~\cite{Bjorken:2009mm,Batell:2009di,Andreas:2012mt} (see also e.g.~\cite{Bergsma:1985qz,Riordan:1987aw,Bjorken:1988as,Bross:1989mp,Blumlein:1990ay,Essig:2010xa,Abrahamyan:2011gv,Boyce:2012ym,Gninenko:2012eq,Adrian:2013zy,Batell:2014mga,Alekhin:2015byh,Dobrich:2015jyk,Banerjee:2016tad,Dolan:2017osp,Berlin:2018pwi,Adrian:2018scb,Dobrich:2018ezn,Beacham:2019nyx,Alemany:2019vsk}) and complement searches in non-accelerator experiments at low energy~\cite{Jaeckel:2010ni,Hewett:2012ns,Alexander:2016aln,Irastorza:2018dyq} as well as collider experiments at very high energy  (see, e.g.~\cite{Jaeckel:2012yz,Jaeckel:2015jla,Brivio:2017ije}).
On the more technical side experimental possibilities are also rapidly expanding, as beams with powers reaching into the megawatt range become available~\cite{Shiltsev:2013zma,Shiltsev:2014jpa,Shiltsev:2017mle,Alemany:2019vsk}.

This motivates a survey of the most promising existing and near future proton beams and their use in searches for new light particles, which we present in this note, with a focus on axion-like particles (ALPs).
We provide an updated calculation of the underlying production process of interest, and focus on the available and future beams, considering a variety of simple geometrical configurations. We leave further subtle but possibly important experimental aspects for future detailed studies.

For concreteness we consider ALPs coupled to two photons with the Lagrangian
\be
\label{alplagrangian}
\mathcal{L}=\frac{1}{2}\partial^\mu a \partial_\mu a -\frac{1}{2}m_a^2 a^2 -\frac{1}{4}g_{a\gamma\gamma} a F^{\mu\nu}\tilde{F}_{\mu\nu}\;.
\ee
We consider this case as it is a standard benchmark scenario that has been used to test many other experimental proposals~\cite{Bergsma:1985qz,Riordan:1987aw,Bjorken:1988as,Bross:1989mp,Blumlein:1990ay,Feng:2018noy,Beacham:2019nyx}. Indeed, ALPs are also well-motivated candidates for new light and intermediate mass particles. As potential remnants of $U(1)$ or shift symmetries, their mass can naturally be lighter than the other scales in the theory and similarly their coupling can be suppressed by an underlying new physics scale. As such, they fulfil the desired properties of a mediator to a dark sector or even dark matter~\cite{Dolan:2017osp}.  
Moreover, ALP production may proceed via a particularly simple mechanism, in the form of a coherent proton--nucleus interaction. In the minimal ALP model~\eqref{alplagrangian} we have only two free parameters, the mass $m_{a}$ and the coupling to photons $g_{a}$. These determine the production rate and lifetime of the ALP, via this coherent process, while detection proceeds via the very displaced decay into two photons. 

Now, a common method to calculate the ALP production cross via this mechanism is to apply the so--called equivalent photon approximation (EPA)~\cite{Budnev:1974de,Dobrich:2015jyk}, whereby the production process may be treated in the proton--nucleus centre--of--mass frame via effective fluxes of quasi--real photons emitted from the ultra--relativistic colliding proton and nucleus. However, as we shall discuss below, this approximation is   based on the assumptions of a sufficiently high energy proton beam, as well as that the ALP mass is not too low (roughly this requires $m_a \gtrsim 100$ MeV). Both of these may break down for the case of light ALP production, with potentially rather low proton beam energies. We therefore present here a more precise calculation of the coherent proton--nucleus production process, valid away from the strict high--energy and not too light ALP mass regimes. We find that the effects of this more precise treatment can be significant: as the beam energy is lowered, a na\"{i}ve application of the EPA can vary dramatically from the exact result, while  in the low mass regime, even if the incoming particles are fully relativistic, the impact is again rather large. These results will therefore have an effect on sensitivity calculations of existing and planned experiments, and therefore to aid such future analyses, we provide a publicly available and flexible tool, based on our improved calculation: the \texttt{Alpaca} Monte Carlo (MC) generator.

The outline of this paper is as follows.
In Section~\ref{sec:flux} we briefly summarise the basic elements of the EPA for coherent ALP production process in proton--nucleus collisions. In Section~\ref{sec:applow} we present our updated calculation of this process. In Section~\ref{sec:comp} we compare our new results with the EPA. The availability and key ingredients of the \texttt{Alpaca} MC are presented in Section~\ref{sec:alpaca}. In Section~\ref{sec:results} we perform our survey and provide numerical results for various beam and dump configurations. Finally, in Section~\ref{sec:conc}, we conclude.

\section{ALP production}
\label{sec:alpprod}
Low mass ALPs can be produced in coherent processes, where the electromagnetic field of the whole proton reacts with that of the whole nucleus. The importance of coherent relative to  incoherent production (via the constituents of the nucleus/proton) originates from the fact that the coherent cross section scales with the full charge of the particles squared, in particular the $Z^2$ of the nucleus.
This is usually calculated~\cite{Dobrich:2015jyk} in the so-called equivalent photon approximation (EPA)~\cite{Budnev:1974de}.  However, as we will discuss below, this needs to be modified for very light ALP masses and low proton beam energies. Before doing so, we briefly review the key elements of the original EPA.

\subsection{Equivalent photon approximation}\label{sec:flux}
\begin{figure}
\begin{center}
\includegraphics[width=6cm]{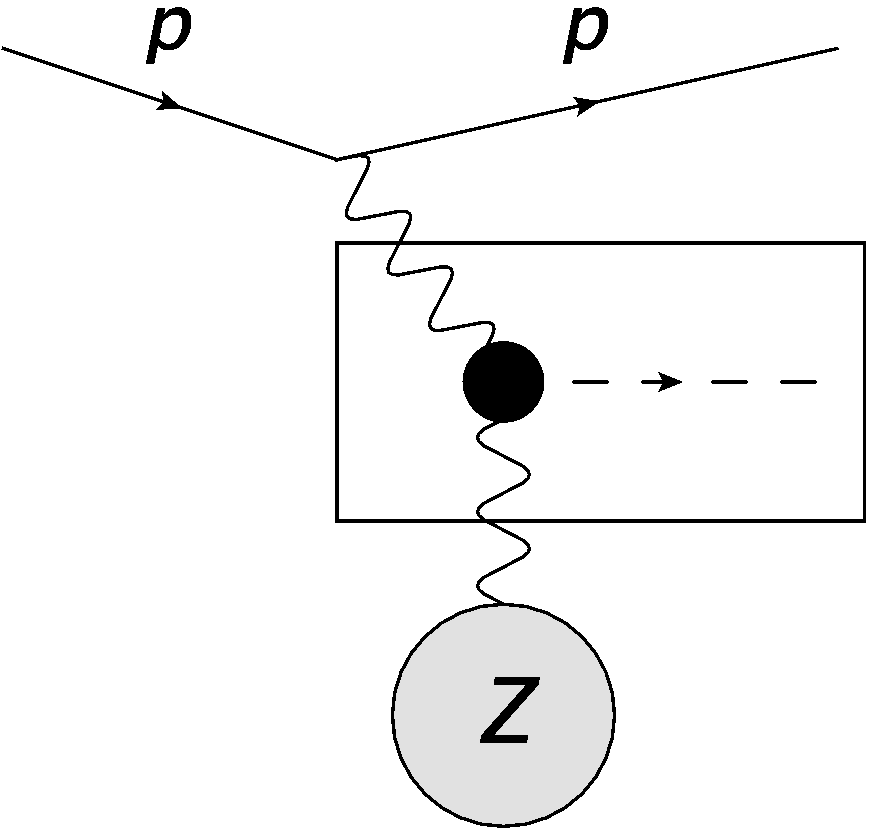}
\caption{Production process of ALPs in coherent proton-nucleus scattering. The partonic cross section for ALP production from two photons is depicted inside the box.}
\label{alpproduction}
\end{center}
\end{figure}

In a proton -- nucleus collision ALPs are produced via the process shown in Fig.~\ref{alpproduction}. In the EPA this is simplified by using distribution functions for the photons carried by the proton and the nucleus. This is completely analogous to the parton distribution functions employed for quarks and gluons in proton proton collisions.
Consequently, we only have to calculate the partonic cross section inside the box of Fig.~\ref{alpproduction} and convolute with the elastic photon distribution functions. 

The flux of quasi-real photons, $n(x)$, determines the distribution of the photons. As shown in e.g.~\cite{Budnev:1974de} this can be quantified in terms of the electric and magnetic form factor of the proton or nucleus,
\begin{equation}\label{WWflux}
N(x,q_\perp^2)=\frac{{\rm d}^2 n(x)}{{\rm d}^2 q_\perp}=\frac{1}{x}\frac{\alpha}{\pi^2}\frac{1}{q_{\perp}^2+x^2 m^2}\left(\frac{q_{\perp}^2}{q_{\perp}^2+x^2 m^2}(1-x)D(Q^2)+\frac{x^2}{2}C(Q^2)\right)\;,
\end{equation}
where $x$ and $q_{\perp}$ are the longtinudinal momentum fraction and transverse momentum of the photon, respectively.
The functions $D$ and $C$ are given in terms of the proton or nucleus electric and magnetic form factors, via
\begin{equation}
C(Q^2)=G_M^2(Q^2)\qquad {\rm and}\qquad D(Q^2)=\frac{4m^2 G_E^2(Q^2)+Q^2_i G_M^2(Q^2)}{4m^2+Q^2}\;,
\end{equation}
where $m$ is the mass of the proton. For each particle the modulus of the photon virtuality, $Q_i^2$, is given by
\begin{equation}\label{qi}
Q_i^2=\frac{q_{\perp,i}^2+x_i^2 m_i^2}{1-x_i} \;,
\end{equation}
where $m_i=m_p(m_N)$ for the proton (nucleus). The kinematic minimum of the virtuality is therefore given by $Q^2_{{\rm min},i}=x_i^2 m_i^2/(1-x_i)$.
The photon virtuality enters the form factors of the proton and the nucleus. For high virtuality the form factors decrease rapidly, reflecting the fact that the composite nature of the particles is resolved. In this region the incoherent contribution (that we do not take into account here) becomes increasingly important.

For the proton we use the dipole approximation
\begin{equation}
G_E^2(Q^2)=\frac{G_M^2(Q^2)}{7.78}=\frac{1}{\left(1+Q^2/0.71 {\rm GeV}^2\right)^4}\;,
\end{equation}
where $G_E$ and $G_M$ are the `Sachs' form factors. 
For a heavy ion of charge $Z$ the electric part is enhanced by a factor of $Z^2$ compared to the magnetic one. We therefore consider only the electric part and write
\be
D(Q^2)=Z^2 F(Q^2)^2\;.
\ee
where $F$ is given by the Fourier transform of the ion proton density $\rho_p(r)$:
\be
F(|\vec{q}|) = \int {\rm d}^3r \, e^{i \vec{q}\cdot \vec{r}}  \rho_p(r)\;,
\ee
in the rest frame of the ion; in this case we have $\vec{q}^2 = Q^2$, so that written covariantly this corresponds to the $F(Q^2)$. For $\rho_p(r)$ we take the Woods--Saxon distribution~\cite{Woods:1954zz}
\be\label{eq:rhop}
\rho_p(r)= \frac{\rho_0}{1+\exp{\left[(r-R)/d\right]}}\;,
\ee
where $d$ is the skin thickness and $R$ is the ion radius. We take~\cite{Chamon:2002mx}
\be\label{eq:rdep}
R  = (1.31 A^{1/3} - 0.84) \, {\rm fm}\;, \qquad d=0.55 \, {\rm fm}\;.
\ee
In the EPA, the cross section is then given by~\cite{Dobrich:2015jyk}
 \be\label{eq:pncross}
\sigma_{pn}^{\rm EPA}=\int {\rm d}x_1 {\rm d}x_2\,{\rm d}^2 q_{1_\perp}{\rm d}^2 q_{2_\perp}\,N_p(x_1,q_{1_\perp}^2)N_n(x_2,q_{2_\perp}^2)\, \sigma(\gamma\gamma \to a)\; .
\ee
where the subprocess cross section is given by
 \be\label{eq:cspart}
 \sigma(\gamma\gamma\to a)=\frac{\pi g_{a\gamma\gamma}^2 m_a}{16}\delta(m_{\gamma\gamma}-m_a)\;.
 \ee
 
\subsection{Going beyond the equivalent photon approximation: low beam energies and low ALP masses}\label{sec:applow}
 
 In deriving the EPA expression \eqref{eq:pncross}, two key approximations are made. Namely, the high energy $s \gg m_N^2$ limit is assumed, i.e. such that the emitting particle is ultra--relativistic, and  the emitted photons are taken to be quasi--on--shell, or more precisely it is assumed that the photon virtuality is much less than the mass of the produced ALP, i.e. $Q^2 \ll m_a^2$.  
 
Both assumptions may be violated in the regions of interest to us. First, some of the beam energies we will consider are relatively low, and deviations from the strict ultra-relativistic approximation may become important.
Second, even for ultra-relativistic beams the assumption that the photon virtuality is small, $Q^2\ll m^{2}_{a}$, may be invalid for the production of relatively light ALPs. Indeed, as the typical photon $Q^2$ is $\mathcal{O}(0.01)~{\rm GeV}^2$, this condition is not fulfilled for light ALPs with $m_a \lesssim 100$ MeV. 

In this section we therefore take into account the full density matrix for the photons at off-shell momenta, as well as providing an exact treatment of the particle kinematics. 
In this way we go beyond the equivalent photon approach, and derive an expression for the ALP production cross section that is valid beyond the ultra--relativistic and low-$Q^2$ regimes. As we shall see, even within this more general setup, for the specific case of pseudoscalar ALP production the result is still rather simple.

We start with the general expression for the ALP production cross section in proton--nucleus collisions~\cite{Budnev:1974de} in the centre--of--mass frame

 \be
 \sigma_{pn} = \frac{\alpha^2}{32\pi^3} \frac{1}{\beta s} \int {\rm d}^3 p_1 {\rm d}^3 p_2 {\rm d}^3 p_a \frac{1}{E_1 E_2 E_a}  \frac{\rho_1^{\mu\mu'}\rho_2^{\nu\nu'} M^*_{\mu'\nu'}M_{\mu\nu}}{q_1^2q_2^2}\delta^{(4)}(q_1+q_2 - p_a)\;.
 \ee
Using
  \be
 {\rm d}^3 p_1 {\rm d}^3 p_2 {\rm d}^3 p_a\delta^{(4)}(q_1+q_2 - p_a)  = \frac{E_1 E_2 E_a}{\tilde{\beta} m_a}  {\rm d}x_1 {\rm d}x_2\,{\rm d}^2 q_{1_\perp}{\rm d}^2 q_{2_\perp}\,\delta(m_a-m_{\gamma\gamma})\;,
 \ee
we find the full expression for ALP production,
 \be\label{eq:crossfull}
 \sigma_{pn} = \frac{\alpha^2}{32\pi^3}\frac{1}{s \beta m_a}\int {\rm d}x_1 {\rm d}x_2\,{\rm d}^2 q_{1_\perp}{\rm d}^2 q_{2_\perp}\,\delta(m_a-m_{\gamma\gamma})\, \frac{1}{\tilde{\beta}}\frac{\rho_1^{\mu\mu'}\rho_2^{\nu\nu'} M^*_{\mu'\nu'}M_{\mu\nu}}{q_1^2q_2^2}\;,
 \ee 
 where
  \be
  \label{suppbeta}
 \beta = \left(1-\frac{2(m_p^2+m_N^2)}{s}+\frac{(m_p^2-m_N^2)^2}{s^2}\right)^{1/2}\;.
 \ee
and the $x_i$ are defined to correspond to the $\pm$ light--cone components of the ALP
\be\label{eq:xdef}
x_{1,2}=\frac{1}{\sqrt{s}}(E_a\pm p_{a,z})=\frac{m_{a_\perp}}{\sqrt{s}}e^{\pm y_a}\;,
\ee
with $m_{a_\perp}=\sqrt{q_{a_\perp}^2+m_a^2}$. As we shall discuss below, this is not necessarily the same definition as is taken when na\"{i}vely calculating with the EPA, and this can have a rather significant effect for general kinematics. Note that this implies
\be\label{eq:x1x2s}
x_1x_2 s= m_{a_\perp}^2\;,
\ee
which we will make use of below. $\tilde{\beta}$ is given in term of the energy $E_i$ and $z$ component $p_{i,z}$ of the outgoing proton and nucleus momenta
\be
\tilde{\beta}=\frac{2 E_1 E_2}{s}\left(\frac{p_{1,z}}{E_1}-\frac{p_{2,z}}{E_2}\right)\;.
\ee
$\rho$ is the density matrix of the virtual photon, which takes the general form
 \be\label{rho}
 \rho_i^{\alpha\beta}=-\left(g^{\alpha\beta}-\frac{q_i^\alpha q_i^\beta}{q_i^2}\right)C_i(q_i^2)-\frac{(2p_i-q_i)^\alpha(2p_i-q_i)^\beta}{q_i^2}D_i(q_i^2)\;,
 \ee
 where the $C,D$ are defined as in Section~\ref{sec:flux}, and $q_i$ is the momentum of the photon emitted from the nucleus $i$, with momentum $p_i$. 
 Finally, $M$ is the amplitude for the $\gamma\gamma \to a$ transition, given by
 \be
 \label{suppmat}
 M_{\alpha\beta}=g_{a\gamma\gamma} \epsilon_{\alpha\beta\sigma\rho}q_1^\sigma q_2^\rho\;.
 \ee
We use~\eqref{eq:crossfull}, together with the supplementary expressions ~\eqref{suppbeta}-\eqref{suppmat} to calculate the ALP production cross section in
\texttt{Alpaca}. We note that to calculate the photon virtuality we do not simply apply \eqref{qi}, which away from the high energy limit is not valid when using \eqref{eq:xdef} to define the momentum fractions $x_i$\footnote{To be precise \eqref{qi} is derived using the Sudakov decomposition $q_1 = x_1 p_1 + \tilde{x}_1 p_2 + q_{1_\perp}$, and equivalently for $q_2$. This definition of the momentum fractions is only equivalent to \eqref{eq:xdef} up to terms of $O(1-\beta)$.}. Rather, this is calculated by using the exactly generated $4$--momenta of the outgoing particles. More precisely, given the ALP 4--momentum calculated in terms of the $x_i$ and $q_{i_\perp}$, the on--shell conditions for the outgoing proton and nucleus are solved numerically to provide the corresponding 4--momenta of the outgoing hadrons. In this way, the kinematically disallowed region, e.g. where in the lab frame we have $E_a>E_{\rm beam}$, is automatically removed by cutting on those events where these on--shell conditions have no solution.

\subsection{Comparison to equivalent photon approximation}\label{sec:comp}

While \eqref{eq:crossfull} corresponds to the result we need, it is useful to consider some further possible approximate simplifications, in particular in order to compare with the result of the EPA. 

We first note that due to the fully antisymmetric form of the matrix element~\eqref{suppmat} all terms $\sim q_{i}$ in the photon density matrix~\eqref{rho} give a vanishing contribution to the cross section. Therefore the scalar (longitudinal) photon polarizations present for off-shell momenta do not contribute at all. In other words, this aspect of the full calculation is in fact consistent with the EPA, where one assumes that the photons are quasi on--shell and therefore that the longitudinal photon polarizations do not contribute. Here, due to the particular form of the $\gamma \gamma \to a$ matrix element this happens to hold true for arbitrary kinematics, see Appendix~\ref{sec:appepa} for further discussion.
 
Continuing with our comparison let us look at a simplified form where we only take into account the (dominant) electromagnetic form factor $D$.
This results in the simple form 
 \be
\rho_1^{\mu\mu'}\rho_2^{\nu\nu'} M^*_{\mu'\nu'}M_{\mu\nu}=4 g_{a\gamma\gamma}^2\frac{\beta^2 s^2}{q_1^2q_2^2}|(q_{1\perp}\times q_{2\perp})|^2 D_1(q_1^2)D_2(q_2^2)\;,
\ee
and the cross section becomes
\begin{align} \nonumber
\sigma_{pn}^{el.}&=\beta\int {\rm d}x_1 {\rm d}x_2\,\,{\rm d}^2 q_{1_\perp}{\rm d}^2 q_{2_\perp}\,\frac{m_{a_\perp}^2}{m_a^2} \frac{(1-x_1)(1-x_2)}{\tilde{\beta}}\,\,\frac{2|(q_{1\perp}\times q_{2\perp})|^2 }{q_{1\perp}^2q_{2\perp}^2}\\ \label{eq:pnp}
&\cdot N_p^{el.}(x_1,q_{1_\perp}^2)N_n(x_2,q_{2_\perp}^2)\,\sigma(\gamma\gamma \to a)\;.
\end{align}
Here, the superscript `el.' denotes that only the electric contribution to the photon flux is included.

How does this compare with the pure EPA result of Eq.~\eqref{eq:pncross}? First, we can see that there is a straightforward factor of $\beta$ which is $\sim 1$ in the high--energy limit, but should be included at lower energies. The factor of $\tilde{\beta}$ is a little more subtle, but one can show that in the high energy limit $\tilde{\beta}\sim (1-x_1)(1-x_2)$, cancelling the corresponding term in the numerator, while again away from this limit this has a non--trivial kinematic dependence. 

A further simplification occurs if we perform angular averaging. We obtain,
\be \label{eq:pnpur}
\sigma_{pn}^{el.} \to \int {\rm d}x_1 {\rm d}x_2\,\,{\rm d}^2 q_{1_\perp}{\rm d}^2 q_{2_\perp}\,\frac{s}{m_a^2}\,\left[x_1 N_p^{el.}(x_1,q_{1_\perp}^2)\right]\left[x_2 N_n(x_2,q_{2_\perp}^2)\right]\,\sigma(\gamma\gamma \to a)\;,
\ee
where we have taken the high energy limit for simplicity, and have factored out the $ \sim 1/x_i$ dependence of the photon fluxes, which simply give a factor of $x_1 x_2 = m_{a_\perp}^2/s$. To compare this with the EPA result,  in \eqref{eq:pncross} 
 we factor out this $1/x_i$ dependence in the same way to give
\begin{align}\label{eq:pncross1}
\sigma_{pn}^{\rm EPA}&=\int {\rm d}x_1 {\rm d}x_2\,{\rm d}^2 q_{1_\perp}{\rm d}^2 q_{2_\perp}\,\frac{s}{s x_1 x_2}\,\left[x_1 N_p^{el.}(x_1,q_{1_\perp}^2)\right]\left[x_2 N_n(x_2,q_{2_\perp}^2)\right]\, \sigma(\gamma\gamma \to a)\; ,\\
&=\int {\rm d}x_1 {\rm d}x_2\,{\rm d}^2 q_{1_\perp}{\rm d}^2 q_{2_\perp}\,\frac{s}{m_{a_\perp}^2}\,\left[x_1 N_p^{el.}(x_1,q_{1_\perp}^2)\right]\left[x_2 N_n(x_2,q_{2_\perp}^2)\right]\, \sigma(\gamma\gamma \to a)\; ,
\end{align}
where in the second line we have inserted the kinematic identity \eqref{eq:x1x2s}. We can therefore see that the dominant contribution to the cross section differs by a factor of $m_{a}^2/m_{a_\perp}^2$, even in the high energy limit. Note that in the limit $Q^2\ll m_a^2$ we precisely have $q_{a_\perp}^2 \ll m_a^2$ and hence this ratio tends to unity. However as discussed above, the condition $Q^2\ll m_a^2$ is generally not satisfied for the production of light ALPs, and so the effect of this may be rather large. 

What is the origin of this additional factor? As discussed further in Appendix~\ref{sec:appepa}, after dropping the longitudinal photon polarization contributions (which are in any case absent for ALP production), in the derivation of the EPA the $Q_i^2 \ll m_a^2$ approximation is applied further to provide a particularly simple formula for generic production processes, as in \eqref{eq:pncross}. It is this approximation which leads to the explicit discrepancy between \eqref{eq:pncross} and \eqref{eq:pnp} for lower ALP masses.

Now, to be precise this extra factor of $m_{a_\perp}^2/m_a^2$ depends on the definition of $x_{i}$ that is applied. Indeed, in the high energy and  $Q^2\ll m_a^2$ regime, there are in principle a variety of choices one could make when defining these, which while equivalent in this limit, may deviate away from it. To investigate this further, we can consider two further approximations:
\begin{align}
x_{1,2} &= \frac{m_a}{\sqrt{s}}e^{\pm y_a}: &{\rm approximation \,1}\\
x_{1,2} &= \frac{\sqrt{k_{a,z}^2+m_a^2}\pm k_{a,z}}{2 E_{1,2}}:& {\rm approximation \,2}
\end{align}
where $k_{a,z}$ is the ALP longitudinal momentum, and $E_{1,2}$ are the energy of the colliding proton (ion), as usual in the c.m.s. frame. This latter definition is used for example in~\cite{Dobrich:2015jyk}, see Eq.~(3.11). In all cases, these definitions are completely equivalent when the $Q^2\ll m_a^2$ (and in the case of approximation 2, the high energy) approximations are imposed, but not otherwise. Note that in both cases in the high energy limit these give 
\be
sx_1 x_2 = m_a^2\;,
\ee
and therefore the factor of $\sim m_{a_\perp}^2/m_a^2$ in \eqref{eq:pncross1} will be absent. However, this is not to say that the result of applying these approximations will necessarily be more precise, lying closer to the exact result. In particular, while this explicit factor is absent, one is implicitly using approximate expressions for the $x_i$ in the photon fluxes, and most importantly in the evaluation of the photon $Q^2$, according to \eqref{qi}; recall that the derivation leading to \eqref{eq:pnpur} relies on the definition \eqref{eq:xdef}. As we will see below, this leads to differences at low $m_a$ which are of the same order as when applying~\eqref{eq:pncross1}. 

\begin{figure}
\begin{center}
\includegraphics[scale=0.85]{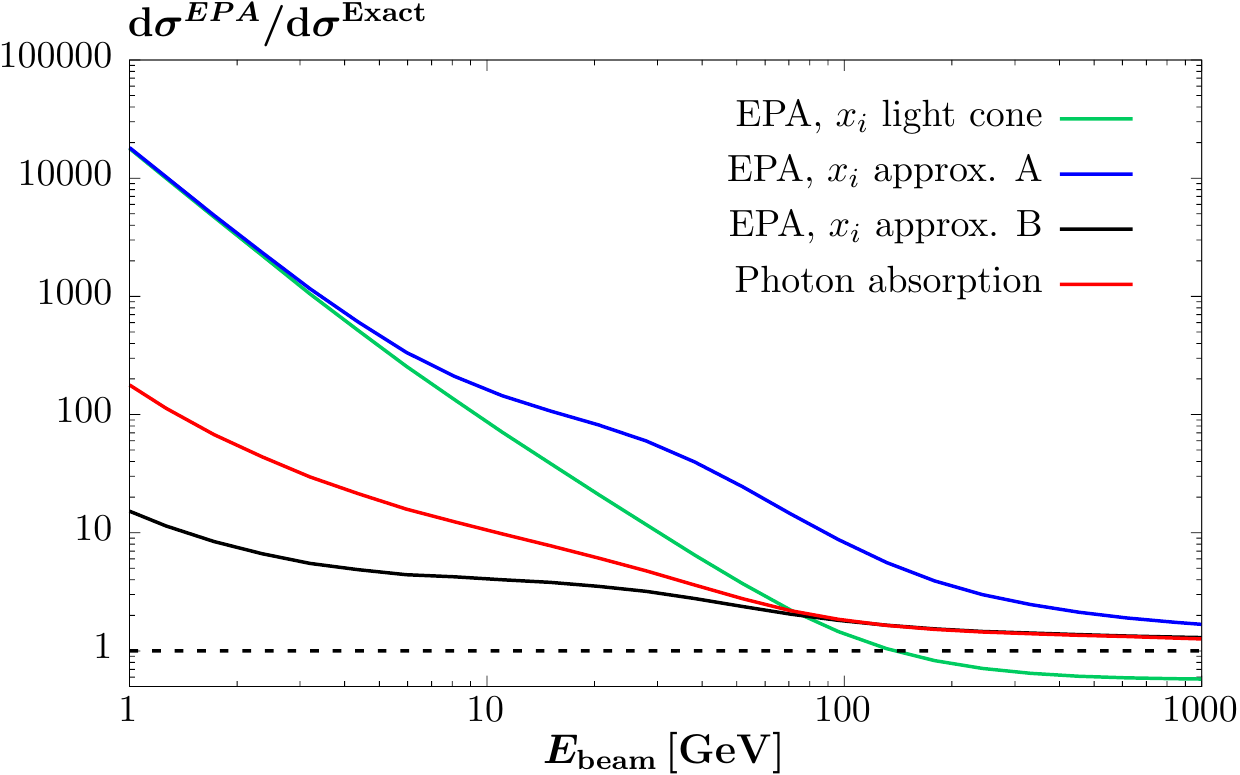}
\caption{The ratio of ALP production cross section calculated using the EPA  \eqref{eq:pncross} and photon absorption~\cite{Dobrich:2015jyk} methods  to the exact \eqref{eq:crossfull} result as a function of the beam energy. For the EPA, a range of definitions of the photon momentum fractions $x_i$ are taken, as defined in the text. A copper target and ALP mass of $m_a = 50$ MeV are taken, while a cut of $\theta < 0.01$ is made on the ALP production angle in the lab frame.}
\label{fig:epaebeam}
\end{center}
\end{figure}

To demonstrate these effects, in Fig.~\ref{fig:epaebeam} we show the ratio of the ALP production cross section calculated using the EPA of~\eqref{eq:pncross} to the full result of~\eqref{eq:crossfull} as function of the beam energy, for an ALP mass $m_a=50$ MeV. For concreteness we consider a copper target and apply a cut of $\theta < 0.01$ on the angle of the produced ALP in the lab frame, without including any further decay effects.  We make these choices to highlight the experimentally more relevant phase space region and isolate any effects due to the ALP decay, as well as to ease the comparison with~\cite{Dobrich:2015jyk}.
For the EPA, we take the light--cone definition \eqref{eq:xdef} as well as the two approximations defined above. We note that taking $x_i = \omega_i/E_i$, where $\omega_i$ ($E_i$) is the photon (parent hadron) energy, for which \eqref{eq:x1x2s} also holds, gives rather similar results to the light--cone definition, and we therefore do not consider it explicitly in what follow. Note that in call cases, energy--momentum conservation is correctly imposed as in \texttt{Alpaca}. We also show the ratio of the cross section calculated in the `photon absorption' approach, directly in the lab frame as discussed in e.g.~\cite{Dobrich:2015jyk}. We can see that for beam energies $E_b \lesssim 100$ GeV the effect of including the full kinematics is dramatic, with the high--energy approximation overestimating the cross section by many orders of magnitude. Moreover, the different definitions of the $x_i$ within the EPA, as well as the photon absorption prediction, all of which  impose the high--energy limit in somewhat different ways, in general lead to rather different results. The remaining deviation at high energy is then due to the ALP mass effects discussed below. The fact that the light cone and approximation 1 results converge at low beam energy can be traced back to the fact that at sufficiently low beam energies larger values of the ALP transverse momentum become kinematically disfavoured, and hence these definitions themselves converge.

\begin{figure}
\begin{center}
\includegraphics[scale=0.85]{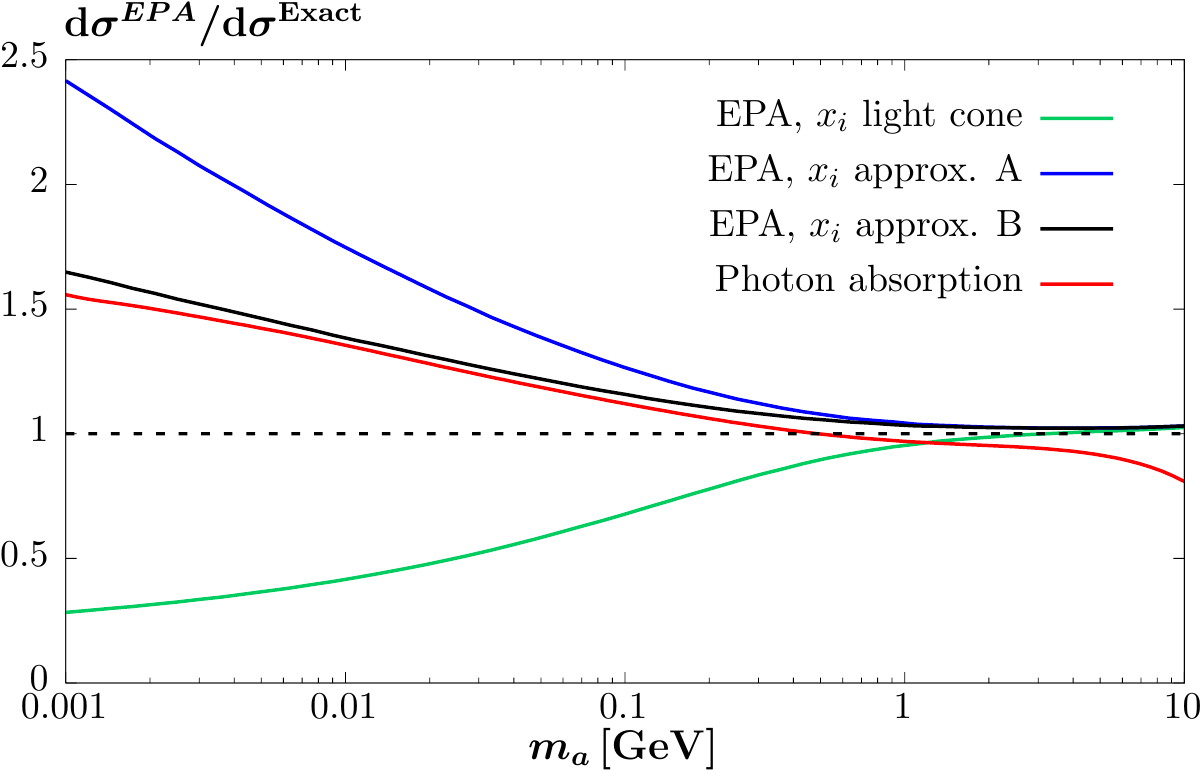}
\caption{The ratio of ALP production cross section calculated using the EPA  \eqref{eq:pncross} and photon absorption~\cite{Dobrich:2015jyk} methods to the exact \eqref{eq:crossfull} result as a function of the ALP mass.  For the EPA, a range of definitions of the photon momentum fractions $x_i$ are taken, as defined in the text. A copper target and beam energy of $E_{\rm beam} = 4000$ GeV are taken, while a cut of $\theta < 0.01$ is made on the ALP production angle in the lab frame.}
\label{fig:epamass}
\end{center}
\end{figure}

We now consider the behaviour as a function of the ALP mass. In Fig.~\ref{fig:epamass} we show the same cross section ratios as above, but as a function of the ALP mass $m_a$. To isolate the mass effects alone, we take a high beam energy $E_b = 4$ TeV to ensure that any deviation from the ultra--relativistic approximation is completely negligible.
We can see that for $m_a \lesssim 0.1$ GeV the results start to differ quite significantly. For the light--cone definition of the $x_i$, we see that the cross section is somewhat suppressed relative to the exact result, in line with the expectations of \eqref{eq:pncross1}, and in particular the factor of $\sim m_a^2/m_{a_\perp}^2$ present\footnote{We note that a closer examination of this cross section, including the kinematic dependence of the photon fluxes themselves, shows that the suppression is in fact only expected to be logarithmic in $\langle m_{a_\perp}^2\rangle/m_a^2$, rather than exhibiting a stronger power--law behaviour, as \eqref{eq:pncross1} might naively suggest.}. For the approximate definitions, as well as the photon absorption cross section, the results are found to be somewhat enhanced relative to the exact prediction. As discussed above, this is due to the imprecise treatment of the kinematics, and in particular the photon $Q^2$, which follows from these treatments. We find that even at high masses there is some residual $\sim \%$ level difference. This is due to the exact calculation of the photon virtuality applied by \texttt{Alpaca}, which in effect includes higher order contributions in $x$ which are neglected in the approximate treatments, as well as the fact that we omitted proton magnetic contribution in the photon absorption case.

\bigskip

In summary, we find rather large differences at lower beam energy and/or ALP mass between the exact result and different applications of the EPA as well as the alternative photon absorption approach, which treats the cross section in the lab frame. Na\"{i}vely, one might assume that the principle cause of the difference is due to the full inclusion of the photon density matrix, including in particular the effect of the off--shell (longitudinal) photon contributions. However, as described above and discussed in more detail in Appendix~\ref{sec:appepa}, the constraining form of the $\gamma\gamma \to a$ amplitude for pseudoscalar ALPs is such that these contributions do not enter, for arbitrary photon virtualities. These differences are therefore purely kinematic in origin, that is due to the precise treatment of the final--state particle kinematics and most importantly the photon virtualities which enter the production cross section.

\subsection{The \texttt{Alpaca} Monte Carlo}
\label{sec:alpaca}

\texttt{Alpaca} is a \texttt{Fortran} based MC generator for ALP production, which applies the results of Sec.~\ref{sec:applow} to calculate the production cross section. The $a\to \gamma\gamma$ decay can then be generated, including full geometric effects from the experimental shielding and decay volume, i.e. the decay length is calculated using the standard expression $l_a = \beta \gamma \tau$, in terms of the lab--frame ALP kinematics, and the decay position is generated according to this. A range of default cuts can then be imposed on the photon energies and their transverse position/separation at the detector. User--defined distributions may be output, as well as unweighted events in the HEPEVT, Les Houches and HEPMC formats. The code and a user manual can be found at: \\
\\
\texttt{http://projects.hepforge.org/alpaca.}

 \section{Results}
 \label{sec:results}

Using \texttt{Alpaca} we can now precisely calculate the ALP production cross section and apply cuts on the decay length and photon separation. This allows us to study the sensitivity
for a number of different experimental setups based on the available beams~\cite{Shiltsev:2013zma,Shiltsev:2014jpa,Shiltsev:2017mle}.
The experimental configuration are essentially modelled after those of the NA62 and SHiP experiments: for lower energies some adaptation of the geometry (smaller shielding etc.) and energy thresholds has been included, but no explicit optimisation has been performed. In particular, we consider the configurations shown in Table~\ref{table:configs}.

\begin{table}[t]
\begin{center}
\renewcommand\arraystretch{1.15}
\resizebox{\textwidth}{!}{
\begin{tabular}{|c|c|c|c|c|c|c|c|c|c|c|}
\hline
Experiment/& shielding & decay & $R_{\rm min}$ [m] & $R_{\rm max}$ [m] &  $d_{\rm min}$ & $E_{\rm ind}$ &Beam  & POT & Target \\
beam	& length [m] &volume [m]& 			& 				& 			&  [GeV]		& energy [GeV] & & material \\
\hline
NA62 (30 days) & 81 & 135 & 0.15 & 1.13 & 0.1  &1 & 400 &$3.9\times 10^{17}$ & Cu\\
\hline
SHiP  & 45 & 60 & 0.15 & 2.5 & 0.1   &1& 400 &$2\times 10^{20}$ & Mo\\
\hline
LHC  & 50 & 50 & 0.15 & 2.5 & 0.1   &1& 7000 &$5\times 10^{16}$& Mo\\
\hline
PSI  & 10 & 10 & 0.15 & 2.5 & 0.1   &0.1& 0.59 & $1.7\times 10^{23}$& Mo\\
\hline
PIP (8 GeV)  & 10 & 10 & 0.15 & 2.5 & 0.1   &0.1& 8 & $7.8\times 10^{20}$& Mo\\
\hline
PIP (120 GeV)  & 50 & 50 & 0.15 & 2.5 & 0.1   &0.5 & 120 & $4.6\times 10^{20}$& Mo\\
\hline
PIP-II (0.8 GeV)  & 10 & 10 & 0.15 & 2.5 & 0.1   &0.1& 0.8 & $1.8\times 10^{21}$& Mo\\
\hline
JPARC RCS  & 10 & 10 & 0.15 & 2.5 & 0.1   &0.1 & 2.95 & $2.5\times 10^{22}$& Mo\\
\hline
JPARC MR  & 10 & 10 & 0.15 & 2.5 & 0.1   &0.5 & 50 & $1.2\times 10^{21}$& Mo\\
\hline\hline
LHC ad. & 50 & 50 & 0 & 2.5 & 0.01  & 1& 7000 &$2\times 10^{17}$& Mo\\
\hline
FCC  & 50 & 50 & 0 & 2.5 & 0.01  & 1& 50000 &$1\times 10^{18}$& Mo\\
\hline
\end{tabular}}
\caption{Experimental configurations for different experiments. The shielding length is the length after the target where all ALPs decaying into photons will be absorbed and cannot be detected. Decays in the ``decay volume" have a chance to reach the detector. The detector is assumed to be cylindrical (i.e. circular) with an inner radius $R_{\rm min}$ and an outer radius $R_{\rm max}$. For good events it will be required that both photons hit the detector with a separation $>d_{\rm min}$ and an individual energy $>E_{\rm ind}$. Finally we have the beam parameters, i.e. the beam energy and the total number of protons on target (POT). Below the double line we show a couple of more aggressive setups for LHC and FCC.}
\label{table:configs}
\end{center}
\end{table}

Following the same procedure as in~\cite{Dobrich:2015jyk} and using the parametrisation of the proton nucleus cross section from~\cite{Carvalho:2003pza} we obtain that the effective integrated luminosity in our beam dump setups is given by
\be
\int \mathcal{L}=\frac{N_{\rm pt}}{53 \,A^{0.77}}\,{\rm mb}^{-1}\,.
\ee
For our sensitivity curves we assume a vanishing background and require $N=3$ events.

\begin{figure}[t]
\begin{center}
\includegraphics[scale=0.60]{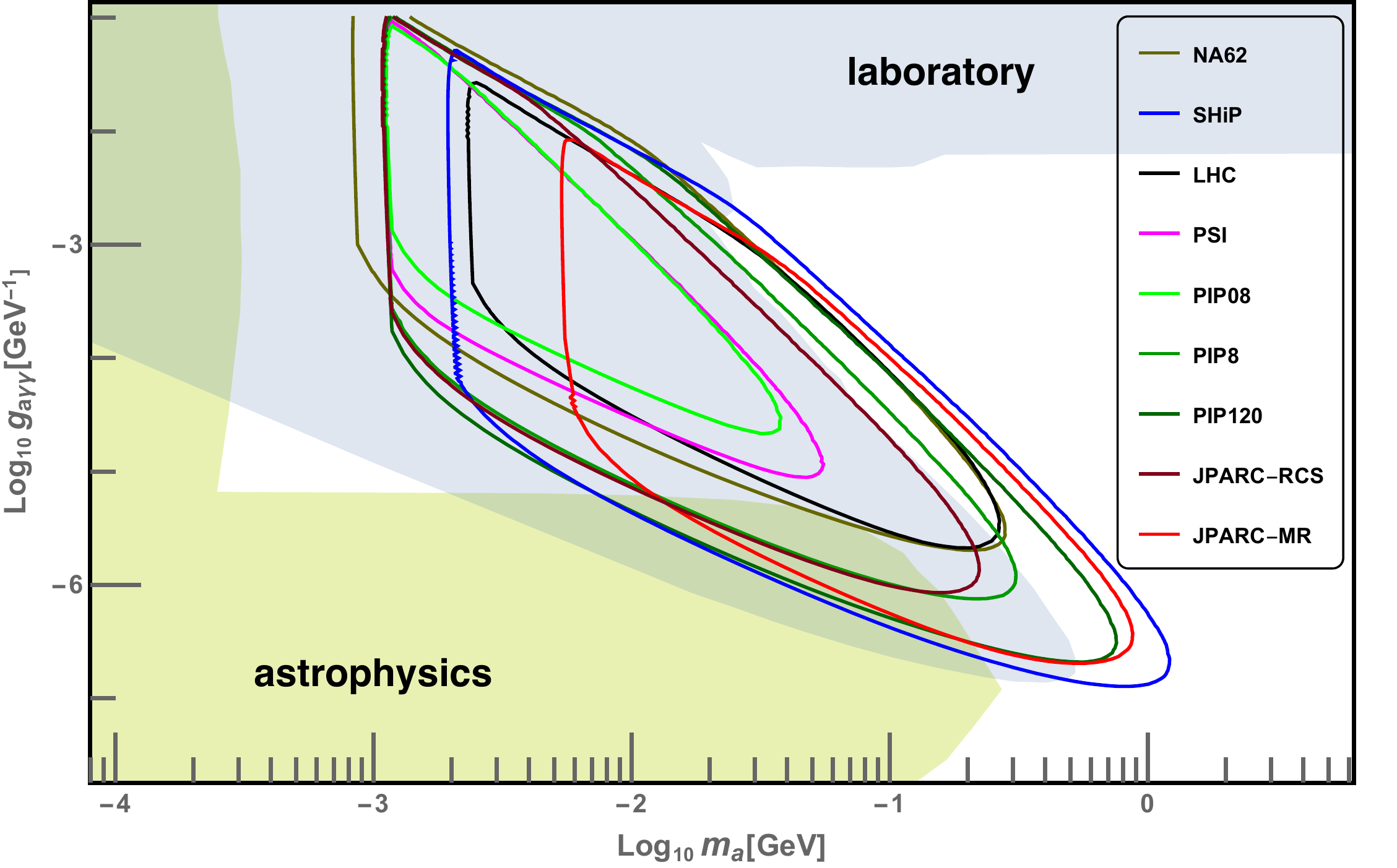}
\caption{Comparison for the sensitivity for the different experimental configurations given in Tab.~\ref{table:configs}. For comparison we also show the recently updated limits from laboratory experiments (grey) and astrophysics (green) from~\cite{Dolan:2017osp}.}
\label{fig:normal}
\end{center}
\end{figure}

The results are shown in Fig.~\ref{fig:normal}. To start with we note that our results for NA62 and SHiP are in qualitative agreement with~\cite{Dobrich:2015jyk} as well the more detailed experimental investigation presented in~\cite{Beacham:2019nyx}. Quantitative differences are expected due to our improved calculation but also due to our simplifying assumptions.

We also note one qualitative difference in all of our curves. At some low mass all considered setups completely loose their sensitivity. This is due to our cut on the {\emph{individual}} photon energy combined with a minimal separation between the photons at the detector. This results from the following combination of effects. 
To achieve a sufficient spatial separation between the photons the angular separation has to be larger than $\sim d_{\rm min}/\ell_{\rm shield}$. The typical angular separation
of the decay products of a boosted ALPs is suppressed by the $\gamma$-factor, $\sim 1/\gamma$. For small masses this is smaller than the minimally required separation.
Therefore, in order to achieve a larger angular separation in the lab frame, low mass ALPs have to decay such that one photon essentially goes in the backward direction.
However, in the lab frame this photon is redshifted and at some point simply does not have enough energy anymore to pass the photon energy cut.
In practise this is not a strong limitation since the affected regions are usually already ruled out by existing limits\footnote{Although one might want to investigate if some of the limits recast from previous experiments might also be affected by this due to a possible limitation in the detectable photon energy.}.

\begin{figure}[t]
\begin{center}
\includegraphics[scale=0.85]{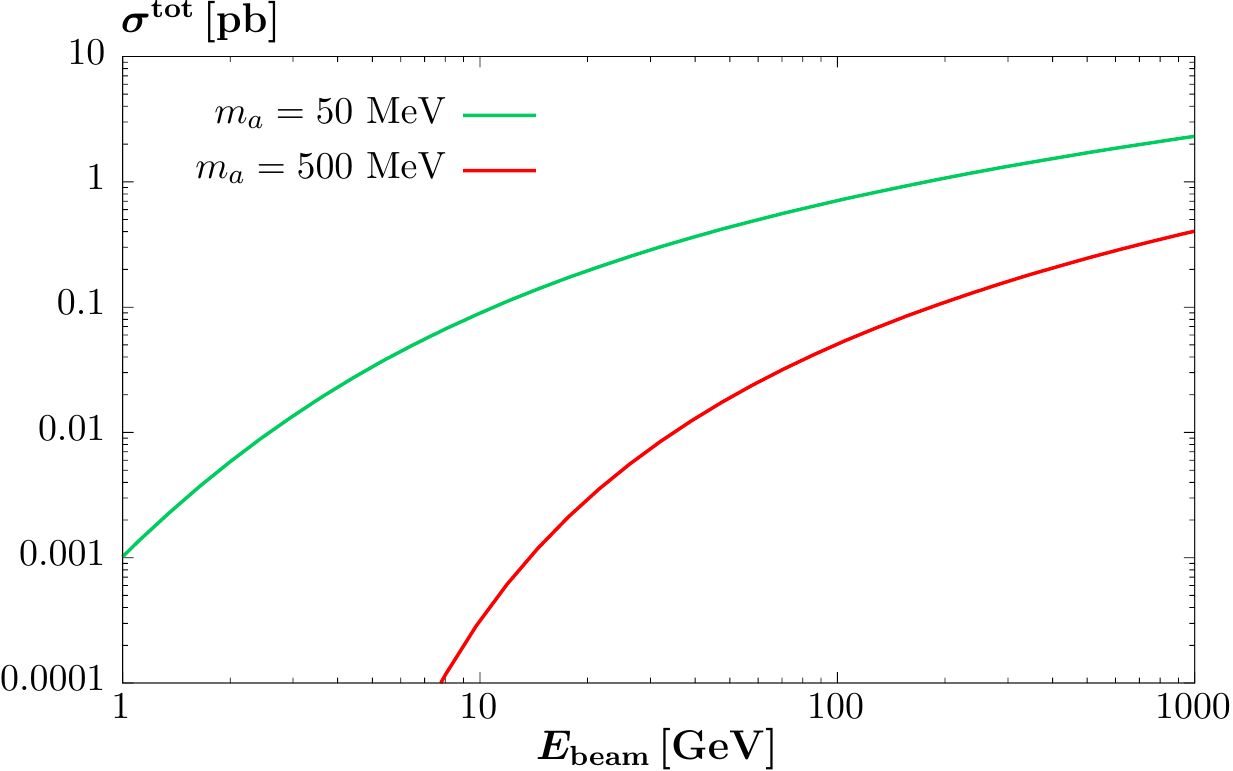}
\caption{Production cross section for ALPs as a function of the beam energy, for two representative choices of ALP mass. }
\label{fig:totalcross}
\end{center}
\end{figure}

\bigskip
Let us now turn to the full range of setups.
From Fig.~\ref{fig:normal} it is clear that the low energy beams do not provide new sensitivity despite their often enormous number of potential protons on target.
There are two main reasons for this. At low masses $m_{a}\lesssim 10\,{\rm MeV}$ the limits are already relatively strong. The only potentially promising target here seems to be the uncovered triangular region that is not yet ruled out by either laboratory experiments or astrophysical considerations. It is however, disfavoured by cosmological arguments~\cite{Millea:2015qra} (not shown in the figure), although these depend on the assumption of standard cosmology. 
Unfortunately, two effects make it difficult for the low energy setups to reach  this region. First, as noted above, requirements on the energy of individual photons lead to a loss of sensitivity in this region. We have also checked a significantly lower energy threshold. While this extends sensitivity to lower masses the sensitivity towards small coupling is not enough. The reason for this is that the total cross section drops rapidly at lower beam energies as shown in Fig.~\ref{fig:totalcross}. Experiments with higher beam energy (such as, e.g., SHiP) therefore have better sensitivity in this region, even if the number of protons on target is significantly lower (provided their photon energy thresholds are sufficiently low).
However, even SHiP with its high number protons on target does not quite reach this region. 

At higher masses $m_{a}\gtrsim 10\,{\rm MeV}$ the most interesting region is the one above the current limits from previous fixed target experiments.
Here, a crucial effect limiting the sensitivity is the decay inside the shielding region. This is mostly dominated by geometry as well as the beam energy.
Keeping the geometry fixed, higher beam energy is advantageous because it leads to higher $\gamma$ factors and therefore longer decay lengths.
Again we can see this in Fig.~\ref{fig:normal}, where NA62, SHiP and the configurations PIP120 and JPARC-MR are most sensitive. The latter shows that high intensity can compensate at least to some degree for a (slightly) lower energy.
Nevertheless the combination of high intensity and high energy provided in SHiP provides the best sensitivity.
We note, however, that geometry is an important factor and some optimisation may be achievable.

\begin{figure}[!t]
\begin{center}
\includegraphics[scale=0.60]{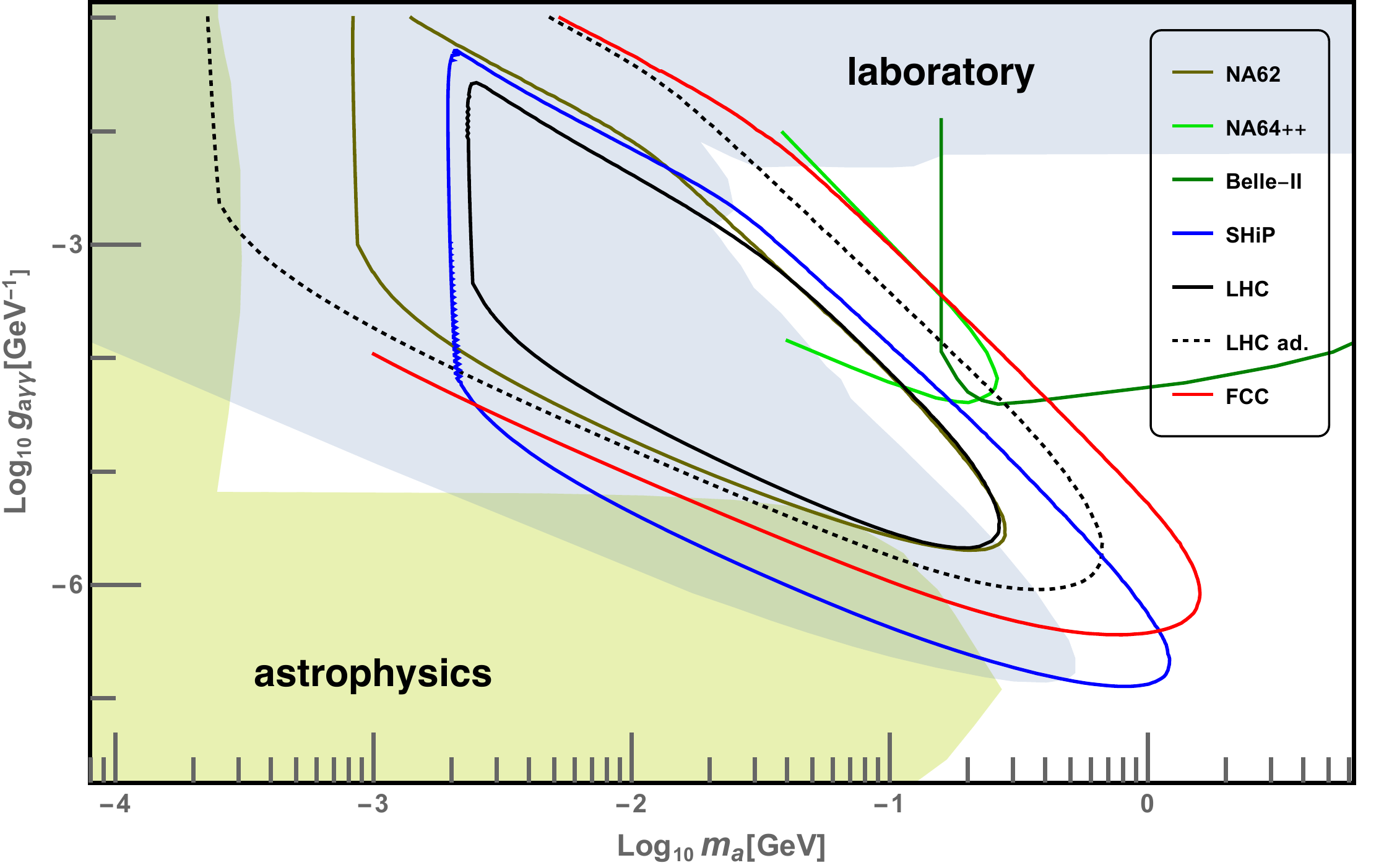}
\caption{Sensitivities for the more ambitious configurations shown below the double line in Tab.~\ref{table:configs}. As in Fig.~\ref{fig:normal} we show limits from laboratory experiments (grey) and astrophysics (green) from~\cite{Dolan:2017osp} as well as the sensitivity of Belle-II~\cite{Dolan:2017osp} and the proposed NA64++ setup~\cite{Beacham:2019nyx}.}
\label{fig:advanced}
\end{center}
\end{figure}

As an interesting special case let us turn to the highest available energies available, namely the LHC beams\footnote{We note that in the case of the LHC the zero background assumption is even more challenging than at SHiP and similar experiments. Amongst other things the high energy leads to a large number of high energy muons
that penetrate large amounts of shielding and are also hard to deflect by magnetic fields. We would like to thank Gaia Lanfranchi for pointing out this issue.}. 
Despite its high energy it can still accumulate a significant number of protons. In the high luminosity phase each fill may contain up to $\sim 5\times 10^{14}$ protons~\cite{Rossi:2130736}.
Assuming about 100 dumps, this gives $5\times 10^{16}$ protons. In the normal configuration outlined in Tab.~\ref{table:configs} this is not sufficient compared to the
dedicated fixed target experiments like SHiP. However, we have also considered a more ambitious option where we use an improved photon resolution such that the photons only need to be separated by $1\,{\rm cm}$, no distance from the center is required  and a higher number of $2\times 10^{17}$ protons is used (numbers below the double line in Tab.~\ref{table:configs}). This is shown in Fig.~\ref{fig:advanced}. As we may expect this shows a significant improvement in the region of larger couplings, as the decay length is increased by the enhanced $\gamma$-factor.

Finally, we have also considered a ``dream'' setup where we used the $50\,{\rm TeV}$ beam of a future FCC~\cite{Benedikt:2018csr}. This allows us to go even further towards closing the gap at large couplings.

\section{Summary and Conclusions}
\label{sec:conc}
In this note we have revisited the coherent production of axion-like particles in proton fixed target experiments. 
We find that a full treatment of the kinematics leads to sizable corrections at small ALP masses $m_{a}\lesssim 100\,{\rm MeV}$.
Even at higher masses the corrections are in the ${\mathcal{O}}(10\%)$ region.
To facilitate the application of this improved calculation we have provided the publicly available {\texttt{Alpaca}} Monte Carlo generator.

As an application we have considered a significant number of proton beams available now or in the near future.
While low energy beams provide the highest intensities, this cannot easily compensate for the increased cross section at higher beam energies.
However, this is dependent on the specific production mechanism and does not fully take into account a potential optimisation that can be achieved if, e.g. the shielding length can be reduced further than we considered.
Overall the SHiP experiment sticks out with its combination of high energy as well as a high number of protons on target, achieving excellent sensitivity.

As a special case we have also considered the LHC and even an FCC at the highest energies. While this requires to overcome enormous challenges in terms of background elimination and very high spatial resolution of the detector, there is potential to extend the reach in the region of relatively large coupling and mass.  

\section*{Acknowledgments} 
We would like to thank Babette D\"obrich and Gaia Lanfranchi for many useful discussions and collaboration on related subjects.
 
 \appendix

\section{Equivalent Photon Approximation: Closer Comparison}\label{sec:appepa}

In this appendix we provide a closer comparison of our results to the equivalent photon approximation, as presented in~\cite{Budnev:1974de}; the notation closely follows this. We start by expanding 
\be
W_{\mu' \nu',\mu\nu} = \frac{\pi}{2 m_{\gamma\gamma}}\delta(m_a-m_{\gamma\gamma})  M^*_{\mu'\nu'}M_{\mu\nu}\;.
\ee
Now, from \eqref{suppmat} we find that the $\gamma\gamma\to a$ amplitude, $M_{\mu\nu}$, for a pseudoscalar ALP has the property that 
\be
q_1^\mu M_{\mu\nu} = q_2^\mu M_{\mu\nu} = 0\;,
\ee
even for off-shell photon momenta $q_{1,2}$, and therefore the amplitudes with scalar photon polarizations are zero for arbitrary momenta. The only contributing terms in the expansion (5.7) of~\cite{Budnev:1974de} are the $W_{TT}$ and spin--flip $W^\tau_{TT}$ terms. We in addition have
\be
W_{+-,+-}=0\;,\qquad W_{++,++}=-W_{++,--}\;,
\ee
where the `$\pm$' indicates the transverse photon polarizations.
Hence
\be
W_{TT} =  - \frac{1}{2}W_{TT}^\tau = \frac{1}{2} W_{++,++}  \;,
\ee
The upshot of this is that (5.7) of ~\cite{Budnev:1974de} becomes
\be
W_{\mu' \nu',\mu\nu} \to \left( R_{\mu \mu'} R_{\nu \nu'}-R_{\mu \nu'} R_{\mu' \nu} \right) W_{++,++} \;,
\ee
where we have used the fact that this will be contracted as in \eqref{eq:pcont} with the momenta $p_i$; that is, we omit the $C$ term on the proton side here, which is not relevant for the discussion which follows.
Here $R$ is the metric tensor in the subspace of $q_{1,2}$:
\be
R^{\mu\nu} = -g^{\mu\nu} + X^{-1}\left[(q_1 q_2)\left(q_1^\mu q_2^\nu + q_1^\nu q_2^\mu\right) - q_1^2 q_2^\mu q_2^\nu - q_2^2 q_1^\mu q_1^\nu \right]\;,
\ee
and $X/m_a^2$ is the square of the photon 3--momentum in the $\gamma\gamma$ c.m.s. frame, given by
\be
X = (q_1 q_2)^2 - q_1^2 q_2^2\;,
\ee
so that in the $q_1^2, q_2^2 \sim 0$ limit we have $X\sim m_a^4/4$. It is therefore no surprise that using the explicit form for $M$ we find
\be\label{eq:mr}
M_{\mu\nu} M_{\mu'\nu'}^* = g_{a\gamma\gamma}^2 X \left( R_{\mu \mu'} R_{\nu \nu'}-R_{\mu \nu'} R_{\mu' \nu} \right) \;,
\ee
consistent with the above results, and with the normalization
\be
W_{++++} =  \frac{\pi}{2 m_{\gamma\gamma}} g_{a\gamma\gamma}^2 X \delta(m_a-m_{\gamma\gamma})\;.
\ee
In terms of this, the quantity
\be\label{eq:sigtt}
\sigma_{TT} \equiv \frac{W_{++++}}{4 \sqrt{X}} = \frac{\pi}{8 m_{\gamma\gamma}} g_{a\gamma\gamma}^2 \sqrt{X} \delta(m_a-m_{\gamma\gamma})\;,
\ee
reduces to $\sigma_{\gamma\gamma}$ \eqref{eq:cspart} in the $q_1^2, q_2^2 \sim 0$ limit.
 Note that
\be\label{eq:pcont}
p_1^\mu p_2^\nu p_1^{\mu'} p_2^{\nu'} M_{\mu\nu} M_{\mu'\nu'}^*  = g_{a\gamma\gamma}^2 \frac{s^2}{4} |q_{1\perp}\times q_{2\perp}|^2\;,
\ee
while defining $\tilde{q}_{i\perp}^\mu \equiv - R^{\mu\nu}p_{i\nu}$ and comparing with \eqref{eq:mr} we find
\be\label{eq:q1q1t}
 |\tilde{q}_{1\perp}\times \tilde{q}_{2\perp}|^2=\frac{s^2}{4 X } |q_{1\perp}\times q_{2\perp}|^2\;.
\ee
As the scalar photon contributions are zero, (5.16) of~\cite{Budnev:1974de} is exact, provided we keep the $\sqrt{X}$ dependence explicit; we will set $\beta,\tilde{\beta}=1$ as we are not interested in corrections to the high energy limit here. We have
\begin{align}\nonumber
\sigma  &= \left(\frac{\alpha}{4\pi^2}\right)^2 \frac{1}{s} \int  {\rm d}x_1 {\rm d}x_2\,{\rm d}^2 q_{1_\perp}{\rm d}^2 q_{2_\perp}\, \frac{2\sqrt{X}}{q_1^2 q_2^2}\left[4 \rho_1^{++}\rho_2^{++}\sigma_{TT}  + 2|\rho_1^{+-}\rho_2^{+-}| \tau_{TT} \cos 2\tilde{\phi}\right]\;,\\ \nonumber
& =  \left(\frac{\alpha}{4\pi^2}\right)^2 \frac{1}{s} \int  {\rm d}x_1 {\rm d}x_2\,{\rm d}^2 q_{1_\perp}{\rm d}^2 q_{2_\perp}\, \frac{16\sqrt{X}}{q_1^2 q_2^2} \rho_1^{++}\rho_2^{++}\sigma_{TT}\sin^2 \tilde{\phi}\;,\\ \label{eq:sigepa1}
& = \frac{\alpha^2}{4\pi^3}\frac{g_{a\gamma\gamma}^2}{s^2}\int {\rm d}Y_X\,{\rm d}^2 q_{1_\perp}{\rm d}^2 q_{2_\perp}\, \frac{X}{q_1^2 q_2^2}\rho_1^{++}\rho_2^{++}\sin^2 \tilde{\phi}\;,
\end{align}
where in the second line we have used that $\tau_{TT} \equiv W_{++,--}/2\sqrt{X} = -2 \sigma_{TT}$,
while
\be
|\rho_{+-}^i| = \rho_{++}^i = 2 D_i \frac{ \tilde{q}_{i\perp}^2}{q_i^2}\;,
\ee
 where the latter holds for $C_i=0$, see (D.3) of~\cite{Budnev:1974de}. Now, if we substitute this into \eqref{eq:sigepa1} and then apply \eqref{eq:q1q1t} we find that all $X$ dependence cancels, and we are indeed left with \eqref{eq:pnp}, as we would expect.
 
 In the EPA approach of~\cite{Budnev:1974de} the $Q_{i}^2 \ll m_{\gamma\gamma}^2$ limit is taken at this point, giving
 \be
  \rho_{++}^i = 2 D_i \frac{ q_{i\perp}^2}{x_i^2 q_i^2}\;,
 \ee
 and $X \sim m_{\gamma\gamma}^4/4$. In fact this essentially just corresponds to substituting this approximate expression for $X$ in \eqref{eq:q1q1t} and elsewhere, and as this dependence cancels, it does not change the result. The only difference is that in the EPA result, in order to write this in the canonical form \eqref{eq:pncross} in terms of the photon fluxes $N_i$, a factor of $x_1 x_2$ is absorbed into their definition, while the approximation $s x_1 x_2 \approx m_a^2$ is applied. Thus there is a factor of $\sim m_a^2/m_{a\perp}^2$ difference between a naive application of the EPA result and the full result, simply due to this approximation being applied in a regime where it does not hold.

\printbibliography

\end{document}